\documentclass[8pt, twocolumn]{article}

\usepackage{cite}
\usepackage{amsmath,amssymb,amsfonts}
\usepackage{algorithmic}
\usepackage{graphicx}
\usepackage{textcomp}
\usepackage{float}
\usepackage{authblk}
\usepackage{geometry}
\title{A Learnable Prior Improves Inverse Tumor Growth Modeling}
\author{Jonas Weidner, Ivan Ezhov, Michal Balcerak, Marie-Christin Metz, Sergey Litvinov, Sebastian Kaltenbach,  Leonhard Feiner, Laurin Lux, Florian Kofler, Jana Lipkova, Jonas Latz, Daniel Rueckert, Bjoern Menze, Benedikt Wiestler }

\begin{document}

\twocolumn[
  \begin{@twocolumnfalse}
    \maketitle
    \begin{abstract}
Biophysical modeling, particularly involving partial differential equations (PDEs), offers significant potential for tailoring disease treatment protocols to individual patients. However, the inverse problem-solving aspect of these models presents a substantial challenge, either due to the high computational requirements of model-based approaches or the limited robustness of deep learning (DL) methods. We propose a novel framework that leverages the unique strengths of both approaches in a synergistic manner. Our method incorporates a DL ensemble for initial parameter estimation, facilitating efficient downstream evolutionary sampling initialized with this DL-based prior. 
We showcase the effectiveness of integrating a rapid deep-learning algorithm with a high-precision evolution strategy in estimating brain tumor cell concentrations from magnetic resonance images. The DL-Prior plays a pivotal role, significantly constraining the effective sampling-parameter space. This reduction results in a fivefold convergence acceleration and a Dice-score of 95\%.
    \end{abstract}
    \vspace{1em} 

    \thanks{
    
J. Weidner, I. Ezhov, L. Feiner und L. Lux are with the Dept. of Computer Science and the Center for Translational Cancer Research, TranslaTUM, Technical University of Munich, Germany (e-mail: j.weidner@tum.de, ivan.ezhov@tum.de, leo.feiner@tum.de, laurin.lux@tum.de).

B. Wiestler and M. Metz are with the Dept. of Neuroradiology, School of Medicine and Health, Technical University of Munich, Germany (e-mail: b.wiestler@tum.de, marie.metz@tum.de)

S. Litvinov and S. Kaltenbach are with the School of Engineering and Applied Sciences, Harvard University, MA, USA (e-mail: slitvinov@seas.harvard.edu, skaltenbach@seas.harvard.edu)

Michal Balcerak and Björn Menze are with the Dept. of Quantitative Biomedicine, University of Zurich, Zurich, Switzerland (e-mail: michal.balcerak@uzh.ch, bjoern.menze@uzh.ch)

F. Kofler is with Helmholtz AI, Helmholtz Zentrum Munich, Germany and with the Center for Translational Cancer Research, TranslaTUM, Technical University of Munich (e-mail: florian.kofler@helmholtz-munich.de)

J. Lipkova is with the Department of Pathology, University of California Irvine, CA, USA (e-mail: jlipkova@hs.uci.edu)

J. Latz is with the Department of Mathematics, University of Manchester, UK (e-mail: jonas.latz@manchester.ac.uk)

D. Rueckert is with the Department of Computing, Biomedical Image Analysis Group, Imperial College London, UK and with the Klinikum rechts der Isar, Technical University Munich, Germany (e-mail: daniel.rueckert@tum.de)

Bjoern Menze and Benedikt Wiestler share senior authorship.

This work was supported in part by the DFG (Deutsche Forschungsgemeinschaft) under Grant 504320104 and by the European High Performance Computing Joint Undertaking (EuroHPC) Grant DCoMEX (956201-H2020-JTI-EuroHPC-2019-1)

Benedikt Wiestler is supported by the NIH (R01CA269948). 

Bjoern Menze and Michal Balcerak acknowledge support by the Helmut-Horten-Foundation.

}
  \end{@twocolumnfalse}
]


 \begin{figure}[H]
    \centering
    \includegraphics[width= \columnwidth]
    {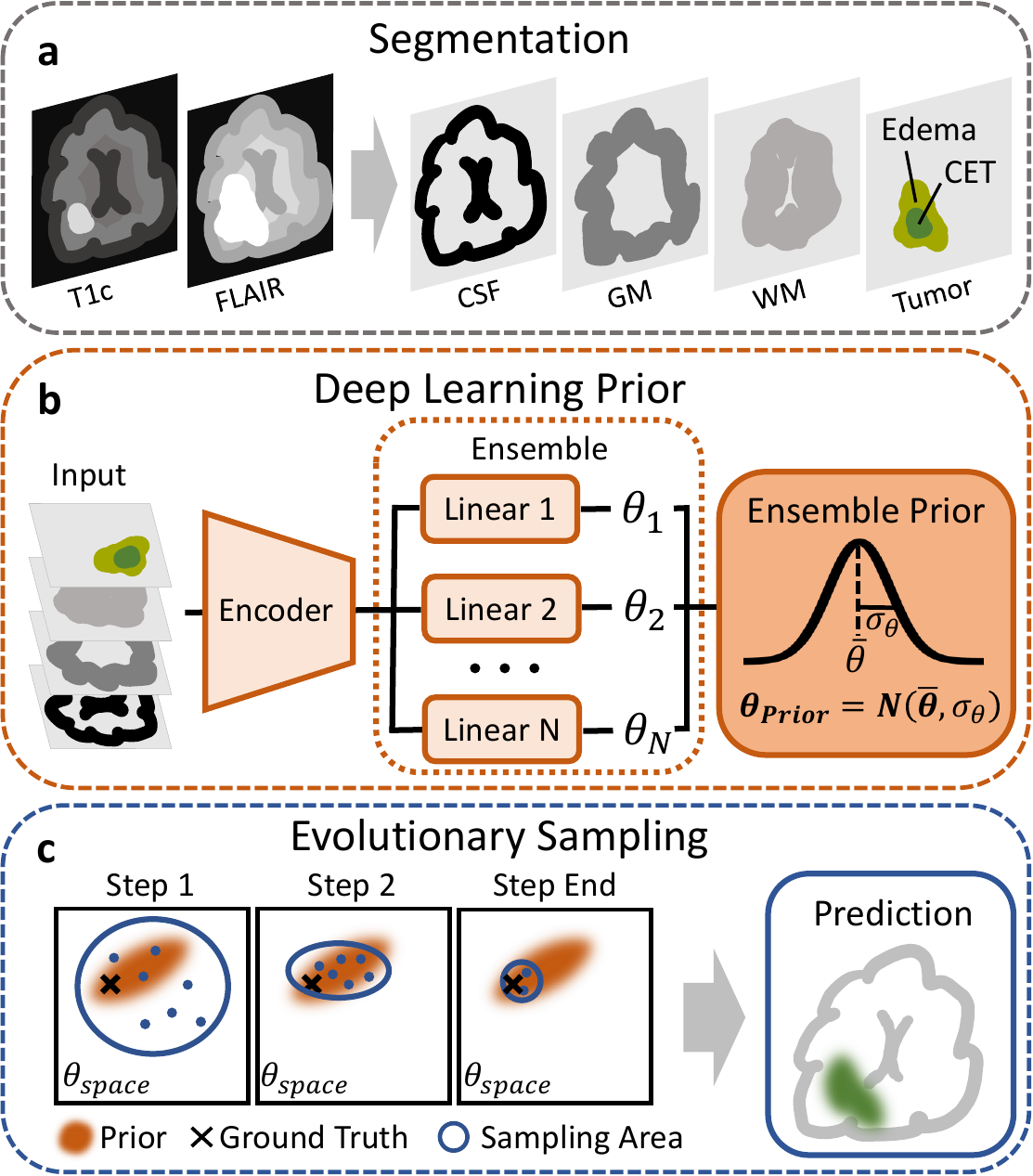}
    \caption{Fitting a tumor model with a combination of DL and evolution strategy. \textbf{(a)} T1c and FLAIR MR images are segmented into white matter (WM), gray matter (GM), and cerebrospinal fluid (CSF). The respective tumor segmentations are combined into one channel. \textbf{(b)} DL provides a fast estimation of the tumor model parameters $\theta = \{x, y, z, \mu_D, \mu_\rho\}$. A Gaussian prior is constructed based on an encoder followed by an ensemble of linear layers. \textbf{(c)} CMA-ES incorporates this DL-Prior and a Dice-score-based likelihood to further optimize the tumor model parameters.}
    \label{fig:overview}
\end{figure}
\section{Introduction}
Glioblastoma is the most common primary malignant brain tumor with a median survival time of only 14-17 months \cite{van2023primary} and a yearly incidence rate of 3.53 per 100,000 \cite{ostrom2020cbtrus}. Despite recent advances in understanding the biology of this disease and prolonged survival with modern combinations of radio- and chemotherapy \cite{weller2021eano}, there is still a tremendous need for improving the treatment of these patients. A central challenge is the tendency of glioblastoma cells to diffusely spread from the primary tumor into the surrounding brain.

Radiotherapy represents a central pillar in brain tumor treatment. The goal of radiation therapy for brain tumors is striking a delicate balance between delivering an effective dose of radiation to the tumor and minimizing damage to surrounding healthy brain tissue. Current glioblastoma radiation therapy planning predominantly relies on MR and CT imaging, with the clinical target volume defined as a uniform 15 mm margin around the resection cavity plus T1 contrast-enhancing tumor \cite{niyazi2023estro}. While this approach aims to address the diffuse spread, it lacks true patient individualization due to its uniformity. Imaging-based individualization of radiotherapy promises to meet the need for more effective therapy.

Radiotherapy planning is limited to the visible boundaries of the tumor on imaging modalities, which is known to only show the ``tip of the iceberg" \cite{niyazi2023estro}. Physical modeling has been explored to predict tumor cell concentrations based on growth patterns to overcome this limitation and uncover the full extent of the tumor. These models aim to incorporate tumor pathophysiology through the spatial and temporal dynamics of the tumor. In the literature, various models are used, ranging from agent-based cellular automaton over multi-scale models to grid-based continuous models, which are applied in this paper \cite{subramanian2022ensemble, jorgensen2023data, hatzikirou2012go, moreira2002cellular, lima2017selection}.

Traditionally, various sampling-based approaches are used to solve the inverse tumor problem, i.e., to estimate the growth model parameters best describing the observed tumor. Simply put, these methods run many forward tumor growth simulations \cite{pabisz2024augmenting}, called samples, in an optimized way to converge toward the most likely set of parameters or towards an entire posterior probability distribution of the parameters \cite{le2015bayesian}. For instance, Bayesian inference executed by Markov-chain Monte Carlo can lead to high-quality model fits \cite{lipkova2019personalized}. With typical forward solvers using the Lattice Boltzmann methods and multi-scope adapted grid methods, a runtime of about 1-3 min per forward run can be achieved for simple tumor models \cite{le2015bayesian, lipkova2019personalized}. However, multiple hundred up to a few thousand samples must be simulated for a reasonable inference result. Thus, although sampling strategies are very precise and, in principle, attractive for clinical use given their convergence robustness, they have a strongly growing complexity with the number of parameters relevant to advanced models. Therefore, these methods are prohibitively costly to calculate, running multiple days per patient \cite{lipkova2019personalized}, forming a significant hurdle for their routine clinical use.

Recently, neural networks have been employed to estimate parameters for brain tumor models \cite{pati2021estimating, ezhov2023learn, zhang2023personalized, zhu2022accelerating} or tumor growth volumes directly \cite{petersen2021continuous}. Convolutional neural networks (CNNs) are designed to analyze images and identify patterns, making them well-suited for analyzing medical images and estimating growth model parameters. Applying CNNs for brain tumor modeling involves training the network to recognize specific features within the MR images associated with tumor growth and progression. After training, these networks can solve the inverse problem of predicting tumor model parameters from an imaging observation to drive a deterministic physics simulation. Despite the fast inference time of less than 5 minutes, the main drawback of a pure DL approach lies in its inflexibility towards modal adaptations and the lack of error guarantees, which can lead to incorrect parameter estimation \cite{grossmann2023can}. This lack of robustness represents a significant challenge for their clinical translation.

Recently, a data-driven model regularized by soft physics-informed penalties demonstrated the ability to predict tumor recurrence better than the current standard procedure \cite{balcerak2023individualizing}. The lack of hard physical constraints within the model presents a limitation, as it may lead to biophysically implausible local minima, which could be addressed by incorporating a physically constrained prior.

The successful integration of deep learning with traditional methods for diverse inverse problems across various fields \cite{xiong2024quantitative, zhao2022learning, jin2024petal} motivated us to explore this approach for brain tumors.

\subsection*{Contributions}
As outlined above, both approaches to solving the inverse problem have individual drawbacks, namely computational cost and lacking robustness. Here, we incorporate the deep learning (DL) approach as a prior for downstream evolutionary sampling. Our methodology combines the speed of DL inference with the reliability and precision of sampling strategies. In more detail,
\begin{itemize}
\item We introduce an ensemble of CNNs to predict tumor growth model parameters in patient space, reducing the mean squared error to simulation ground truth by 80\% compared to existing methods.
 
\item We reduce the effective parameter space for subsequent precise evolutionary sampling, based on the DL prediction, effectively accelerating convergence time by a factor of five without sacrificing precision.
\end{itemize}

\section{Methods}
\subsection{Framework Overview}
Our goal is to predict the tumor concentration for each voxel, which is not visible in conventional imaging modalities, based on physically constrained modeling (Figure \ref{fig:overview}). As input to our model, contrast-enhanced T1 (T1c) and FLAIR MRI scans are used, highlighting higher and lower tumor concentrations, respectively (Figure \ref{fig:overview} a). In a preprocessing step, the tumor regions are segmented in both scans into the subregions contrast-enhancing tumor (CET) and edema, using BraTS Toolkit \cite{kofler2020brats}. The tissue segmentation is generated based on the T1c scan using the S3 software \cite{lipkova2019personalized}. 

The tumor and tissue segmentations are fed into a neural network to predict the physical tumor model parameters (Figure \ref{fig:overview} b). The premise is that the network learns the growth dynamics based on the underlying tissue. Therefore, an overlay of both tumor segmentations is fed into the network as a single channel(edema: 0.33, CET:0.66), whereas the other input channels contain the patient-specific tissue information: white matter (WM), gray matter (GM) and cerebrospinal fluid (CSF) to facilitate patient-specific modeling. The tissue information in each channel is continuous and can be interpreted as the ratio of the given tissue type per voxel. The network output consists of scalar parameters that determine the physical process of tumor development. From those parameters, it is possible to simulate the growth process of the tumor within the patient's brain. 

We discuss the network architecture in section \ref{sec:methDL}. The DL-Prior is included in subsequent evolutionary sampling, which is explained in section \ref{sec:methSample}. Evolutionary sampling is a computational technique that leverages multiple forward simulations to converge toward the optimal parameters governing tumor behavior. The DL-Prior plays a crucial role in guiding the exploration of this parameter space.

\subsection{Tumor Growth Model}
The tumor progression can be distinguished into two main parts: growth and diffusion. From those two phenomena, the tumor concentration $c$ can be modeled by the following partial differential equation:

\begin{equation}
\frac{\partial c}{\partial t } = \nabla \cdot (D\nabla c) + \rho c (1-c) 
\end{equation}
The first term describes the diffusion process with diffusion coefficient $D$, and the second term describes the growth process over time $t$. In the case of the applied Fisher-Kolmogorov equation \cite{harpold2007evolution}, a logistic growth process with a growth parameter $\rho$ is assumed.

\subsection{Tumor Simulation}

For the simulation, three types of brain tissue are distinguished: WM, GM and CSF. For each voxel $i$ the probability $p$ of all tissues adds up to one: $p_{_{WM}} + p_{_{GM}} + p_{_{CSF}} = 1$.

The diffusion coefficient is assumed to differ for each tissue type \cite{lipkova2019personalized}. The ratio of diffusion within WM and GM is fixed to $D_{_{GM}} = \frac{1}{10} D_{_{WM}}$, whereas no diffusion is allowed inside CSF ($D_{_{CSF}} = 0$), as the tumor physiologically cannot grow there.

Additional parameters are the tumor origin $\{ x, y, z\}$ and the total time of tumor growth $T$. We select a single tumor origin as the starting point of the simulation. This simulated origin cannot necessarily be directly associated with the actual tumor-originating cell in medical cases. Thus, the original set of parameters for the forward solver is $\theta_{orig} = \{ x, y, z, \rho, D_{_{WM}}, T \} $. As only one time-point is realistically available for radiotherapy planning, $\theta_{orig}$ is ambiguous: for example, a faster-growing tumor over a shorter time and a slower-growing tumor over a longer time are indistinguishable based on a single time point. Therefore we introduce end-time independent parameters $\mu_D = \sqrt{D_{_{WM}}T}$ [cm] and $\mu_\rho= \sqrt{\rho T}$ [unitless] leading to the reduced parameter set: $\theta = \{ x, y, z, \mu_D, \mu_\rho \}$. The 3D tumor cell density $c$ is a function of these parameters. For simulation, the GliomaSolver \cite{lipkova2019personalized} software is used.
 
\subsection{Data}
For the network training, 30,000 simulations are created using 140 real patient tissues from the BraTS dataset \cite{menze2014multimodal} to simulate synthetic tumors. By explicitly incorporating tissues from BraTS patients, whose anatomy is already altered by tumors, we introduce pathological anatomy variations that the network is likely to encounter in real patient data. This approach aims to make the model invariant to tumor-induced anatomical differences and typical tissue deformations, thereby enhancing its ability to generalize to diverse patient data. The dataset is split into 28000 training samples and 2000 validation samples.  

For the additional test set for extensive validation of all methods, including costly additional sampling, 180 simulations are conducted in another 140 BraTS patient anatomies with random parameters.

For validation, we apply all methods to preoperative imaging data from nine real glioblastoma patients to compare their ability to fit preoperative tumor segmentations instead of the comparison to synthetic ground truth.

\subsection{Deep Learning Network}
\label{sec:methDL}
\subsubsection{Network Training}
The network encoder (Figure \ref{fig:overview} b) is based on a residual neural network (ResNet) \cite{resnet}. In contrast to former experiments \cite{ezhov2023learn}, the network input consists of four channels with dimensions $128 \times 128 \times 128$, concatenating the three patient-specific tissue masks and the tumor segmentation map. All other parameters and conditions are maintained as in the prior experiments \cite{ezhov2023learn}. The network outputs are the five model parameters $\theta$ optimizing an $L_2$ loss. The simulation parameters are sampled based on clinical experience in the same way as introduced in the Learn-Morph-Infer (LMI) method \cite{ezhov2023learn}.

For the original model, the input tumor is generated by random thresholding of the synthetic ground truth tumor at values $c^{th}_{Edema} \in [0.05, 0.5]$ and $c^{th}_{CET} \in [0.5, 0.85]$. All voxels containing ground truth tumor concentrations above these thresholds are considered edema ($c^{th}_{Edema} \leq c < c^{th}_{CET}$) or CET ($c \ge c^{th}_{CET}$) segmentations for training.

The training was conducted on an NVIDIA Quadro RTX 8000 with 50GB RAM. We trained for 85 hours over 50 epochs. 

\subsubsection{Network Ensemble}
An ensemble of ten networks is introduced to increase the model robustness towards multiple edema and CET thresholds. The last layer of the network ensemble is fine-tuned for a certain threshold ($c^{th}_{Edema}, c^{th}_{CET}$) in the ground truth training data. The premise is that the ensemble mean is more stable toward single network failure, and the ensemble standard deviation can quantify the uncertainty of the prediction. Technically, we fixed all weights except the last linear layer and retrained it on different thresholds. A Gaussian prior is generated from this equally weighted ensemble (see section \ref{sec:prior}).

\subsection{Evolutionary Sampling}
\label{sec:methSample}
Introducing a prior drastically reduces the effective parameter space for the subsequent sampling strategy. The following sections describe the integration of this prior into a covariance matrix adaptation evolutionary strategy (CMA-ES) \cite{hansen2001completely, hansen2003reducing}. 

\subsubsection{Prior}
\label{sec:prior}
A Gaussian distribution around the DL-ensemble estimated parameters is defined as prior. The prior distribution is based on the ensemble mean ($\bar{\theta}^i_{dl}$) and its standard deviation ($\sigma^i_{dl}$) for each parameter $i \in \theta$. It is assumed that the individual parameters follow independent Gaussian distributions.

\begin{equation}
P(\theta|\bar{\theta}_{dl},\sigma_{dl}) \sim \prod_{i} \frac{1}{\sigma^i_{dl}\sqrt{2\pi}} \exp \left(-\frac{(\theta^i- \bar{\theta}^i_{dl} )^2}{2 (k \sigma^i_{dl})^2 } \right)
\end{equation}

The scaling parameter $k$ is introduced to relax or tighten the variance of the prior based on empirical findings. It is found that the model sampling performs better if we relax the prior contribution compared to the likelihood. This is done by increasing $k$ to a value of 10. In a strict Bayesian formulation the prior does not depend on the data. As patient data is used for DL inference, the prior does depend on the data, so technically, we apply empirical Bayes \cite{casella1985introduction}.

\subsubsection{Likelihood}
The likelihood describes the probability of observing the MRI tumor segmentations (data $\mathcal{D} = \{Y_\text{CET}, Y_\text{Edema}\}$) given a set of tumor model parameters $\theta$. It is composed of the likelihoods of edema and CET where $c^{th} = \{c^{th}_{CET}, c^{th}_{Edema}\}$. 

\begin{align}
P(\mathcal{D}|\theta, c^{th}) =   P(Y_{\text{CET}}|\theta, c^{th}_{CET}) \cdot P(Y_{\text{Edema}}|\theta, c^{th}_{Edema})   
\label{eq:likelihood}
\end{align}

The respective probability of observing the segmentations with a given set of parameters is assumed to be described by the Dice-score (\text{DSC})\footnote{Approximating the likelihood function with the Dice-score is mathematically inaccurate. For example, a vanishing Dice-score is not equivalent to a zero probability of observing the data. Nevertheless, the real likelihood function is unknown, and modeling it would imply further assumptions\cite{lipkova2019personalized} that we want to avoid.} \cite{dice1945measures} between the segmentation and the proposed concentration above the threshold $c^{th}_{Edema}$, and equivalently for T1c $c^{th}_{CET}$. 

\begin{equation}
P(Y_{\text{Edema}}|\theta, c^{th}_{Edema}) = \text{DSC} ( Y_{\text{Edema}}, c(\theta) > c^{th}_{Edema} ) 
\end{equation}

\subsubsection{Posterior}
The posterior is calculated with the DSC-based likelihood and the DL-based prior. 

\begin{equation}
P(\theta|\mathcal{D}) \sim P(\mathcal{D}|\theta, c^{th}) \cdot P(\theta|\bar{\theta}_{dl},\sigma_{dl})
\label{eq:posterior}
\end{equation}

The maximum a posteriori probability (MAP) estimate provides the most likely tumor model parameters $\hat{\theta}_{\text{MAP}}$ under the described assumptions. 
\begin{align}
\hat{\theta}_{\text{MAP}} = &\arg\max_{\theta} P(\theta , c^{th}| \mathcal{D}) \\
 =  &\arg\max_{\theta} \int_{c^{th}} P(\theta , c^{th}| \mathcal{D})  dc^{th} \\
  \overset{\text{(\ref{eq:posterior})}}{=}  &\arg\max_{\theta} \int_{c^{th}} P(\mathcal{D}|\theta, c^{th})  dc^{th} \cdot P(\theta|\bar{\theta}_{dl},\sigma_{dl}) \\
 \overset{\text{*}}{\approx} &\arg\max_{\theta} P(\mathcal{D}|\theta, \overline{c^{th}}) \cdot P(\theta|\bar{\theta}_{dl},\sigma_{dl})
 \label{eq:MAP}
\end{align}

We applied the midpoint approximation (*) over the likelihood for the average thresholds $\overline {c^{th}} = \{0.25, 0.675\} $, as the threshold values are sampled over a uniform distribution, and the difference in likelihood is small on a relative scale (about max 5\%)  for various thresholds. As there is no analytical solution for $\hat{\theta}_{\text{MAP}}$, the optimization is carried out by CMA-ES as described in the following.  

\subsubsection{Sampling Algorithm}
The novelty of our approach lies in integrating a DL-Prior into sampling strategies. This is realized by the state-of-the-art covariance matrix adaptation evolution strategy algorithm \cite{hansen2001completely} (Figure \ref{fig:overview} c). 

The CMA-ES algorithm is employed due to the inherently ill-posed, non-convex, and non-differentiable nature of the inverse tumor growth problem \cite{gholami2016inverse}. Furthermore, CMA-ES exhibits robust performance on unsteady problems frequently encountered at tissue interfaces. For instance, a tumor might not grow at all if its origin lies in a voxel where CSF is dominant, whereas it may proliferate unimpeded in an adjacent white matter voxel.

The idea of the CMA-ES is based on sampling within a multivariate Gaussian distribution in parameter space. In each step, samples are evaluated according to their fitness. In our case, the fitness function is represented by the estimated posterior $P(\theta | \mathcal{D})$, describing the probability of the parameter set $\theta$ given the measured data $\mathcal{D}$ (Eq. \ref{eq:posterior}). The fittest samples are selected, and the sampling mean and the covariance matrix of the multivariate Gaussian are adapted accordingly \cite{auger2012tutorial}. Eventually, the sampling distribution converges to the MAP $\hat{\theta}_{\text{MAP}}$ (Eq. \ref{eq:MAP}). We sampled for a maximum of 600 simulations. 

To ensure an equitable comparison of runtime between sampling methods with and without a prior, we implement a specific initialization protocol for the absence of a prior, the naive sampling. The tumor origin is initialized at the center of mass of the FLAIR segmentation. The diffusion coefficient $\mu_D$ and $\mu_{rho}$ are initialized at its lower limit, given that the forward simulation runtime increases linearly with $\mu_D$. 
This initialization results in a sampling strategy that converges towards optimal parameters while commencing with a small tumor and subsequently increasing in size. As smaller tumors facilitate expedited simulation time, this initialization improves the entire sampling runtime. For this comparative analysis, a subset of 13 patients was randomly selected for cost reasons. We sampled until a convergence up to 99 \% of the maximum posterior up to a total of 2000 samples for the case without prior and 600 samples for the case with DL-Prior. All simulations were conducted on an AMD EPYC 7313 16-core CPU. We published our code on GitHub\footnote{github.com/jonasw247/a-learnable-prior-improves-inverse-tumor-growth-modeling}.

\subsection{Evaluation}
The predicted parameters can be compared to the ground truth as an intermediate evaluation. Here, the distance to the ground truth origin and the deviation from the ground truth growth and diffusion parameters are of interest. Note, however, that due to the ill-posedness of the problem, several combinations of parameters can describe a given tumor observation equally well. In view of the potential clinical application for individualized radiotherapy, the accurate estimation of tumor cell concentration is more important.

Thus, different metrics are applied to evaluate the similarity between predictions and ground truth concentration. The mean absolute error (MAE) and mean squared error (MSE) provide information about the pixel-wise differences between the proposed tumor and the ground truth. The problem with those metrics lies in the averaging over all pixels. For example, a tumor where every voxel concentration differs a bit can have the same error as a tumor where some areas are utterly distinct while some regions fit very well. Therefore, the commonly used Dice-score \cite{dice1945measures} is also applied. The Dice-score measures the overlap of two binary segmentations. A tumor density threshold must be applied as the solver's output is continuous. 

For synthetic data, where ground truth tumor concentration is available, any clinically relevant threshold can be selected to binarize the ground truth and the proposed tumor density map. As a benchmark, the atlas-specific network LMI is integrated into our synthetic workflow \cite{ezhov2023learn}.

To ensure statistical robustness, we have highlighted the significant results by marking them with an asterisk (*). We set the threshold for significance to a p-value of 1\%. If not stated otherwise, we compared the different methods to our proposed method, "DL-Prior + Sampling". As we cannot assume Gaussian distributions and as we analyze paired data, we used the Wilcoxon signed-rank test.

\section{Results}
\begin{figure*}
    \centering
    \includegraphics[width=\textwidth]{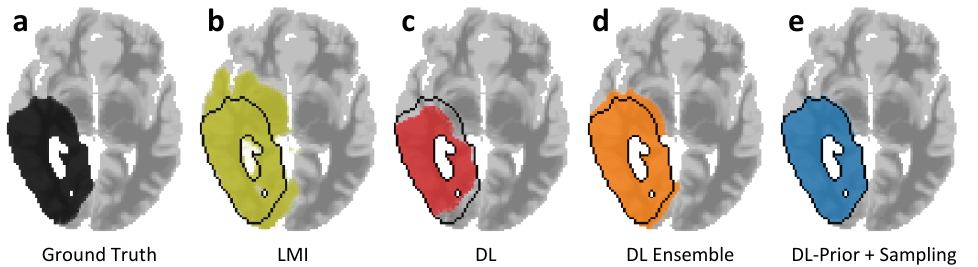}
    \caption{Example tumors created by different models are compared. The ground truth tumor is shown in black (a). Its outline is displayed in the other images as well. The atlas-based LMI approach (b), our DL-Prior (c), the ensemble mean (d), and the combination of the DL-Prior and sampling (e) are compared.}
    \label{fig:example}
\end{figure*}

\begin{figure}
    \centering
    \includegraphics[width=\columnwidth]{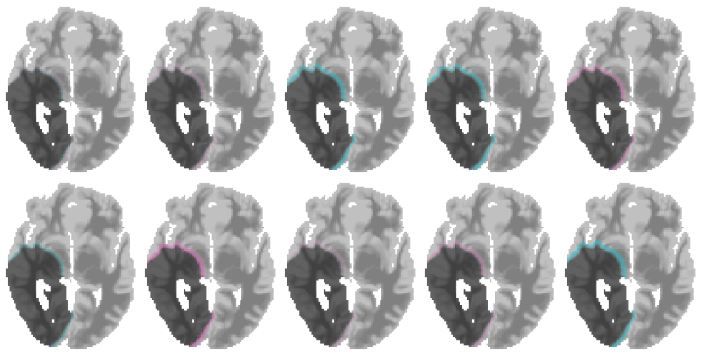}
    \caption{Example simulations with parameters created by the ensemble of the networks. The difference (cyan, positive; pink, negative) to the ensemble mean (gray, Figure \ref{fig:example} d) is shown.}
    \label{fig:ensemble}
\end{figure}

In this section, we present the evaluation of various methodologies to predict tumor cell distribution. We compare the following methods:
\begin{itemize}
    \item \textbf{LMI}, which is a DL method requiring atlas registration \cite{ezhov2023learn}. 
    \item \textbf{DL}, which is the in-patient-space DL approach with a single network.
    \item \textbf{DL Ensemble}, which is the mean of an ensemble of networks, finetuned for different Edema and CET thresholds.
    \item \textbf{Naive Sampling}, which is the CMA-ES.
    \item \textbf{DL-Prior + Sampling}, which is the subsequent evolutionary sampling incorporating a DL-Prior based on the DL ensemble.
\end{itemize}
First, we qualitatively compare the proposed tumor concentrations, highlighted in Figure \ref{fig:example}. Subsequently, we delve into a detailed comparison of estimated parameters. This is illustrated in Table \ref{table:parameterResults}, where deviations from ground truth parameters are analyzed. The following evaluation of tumor concentrations, as detailed in Figure \ref{fig:diceResults} and Table \ref{table:concentraionResults}, provides the most relevant metric of different methodologies. Additionally, we compare the runtime efficiency of these methods in Figure \ref{fig:convTime}, evaluating the convergence times of each approach. Finally, we evaluate our method with nine real patients (Figure \ref{fig:realPatients}, Table \ref{tab:realPatients}).

\subsection{Deep Learning Methods}
In the first results part, we focus on the DL methods described in Figure \ref{fig:overview} b.

\subsubsection{Qualitative Results}
In Figure \ref{fig:example}, a qualitative comparison of the different approaches is visualized. It is visible that the LMI tumor prediction (b) does not align perfectly with the underlying tissue. This discrepancy is caused by the registration process from the atlas space into the patient space. Furthermore, the LMI prediction overestimates the tumor size. Our in-patient space method (c) slightly underestimates the tumor size, whereas the ensemble mean (d) overestimates it. 

\subsubsection{Estimated Parameter Comparison}
The parameter estimation is compared to the simulated ground truth (Table \ref{table:parameterResults}). All parameter estimations are Gaussian distributed, as expected\footnote{The distribution of origin-deviation is not Gaussian itself, but it is composed of three Gaussian distributions (x,y,z).}. Therefore, only the mean deviation and the RMSD are reported.

The estimation of the tumor origin is generally precise, with an average deviation of 3.7 mm to the ground truth for the single network (3.6 mm for the ensemble mean). This is twice as good as the LMI approach, with a mean displacement of 8.7 mm. As a baseline, the edema segmentation's center of mass is included as a rough estimation. The CET center of mass mean displacement is 12.7 mm away from the ground truth, demonstrating the superiority of our method.

For $\mu_D$, the time-independent diffusion parameter, a three-fold improvement in root means square deviation (RMSD) is found for DL over LMI. An average overestimation (16.5 mm) can be found for LMI, whereas our approach underestimates $\mu_D$ slightly (-2.4 mm). Regarding the unitless, time-independent growth parameter $\mu_\rho$, a two-fold improvement (in RMSD) is found, whereas no systematic deviation is detected.

The DL ensemble (orange), finetuned with different edema and CET thresholds, performs similarly to our original (DL, non-ensemble) model (red) on parameter level.

\begin{table*}
\centering
\caption{The tumor model parameter mean displacement (MD) and the root-means-squared deviation (RMSD) from the simulation ground truth are shown for different methods. MD is a criterion for the systematic deviation, the shift between the means. RMSD is a measure of the spread of differences in distribution and, thus, determines the quality of the fit. The distance from the edema and CET center of mass to the ground truth tumor origin is included. A significant improvement in origin estimation of DL-Prior + sampling is found over the core center of mass. For $\mu_D$ and $\mu_\rho$, subsequent sampling results in worse parameter results. This highlights that the problem is ill-posed, meaning that finding a better fit to the observed tumor data can result in deviations or instability in the estimated parameter values (see also Table \ref{table:concentraionResults})}

\label{table:parameterResults}
\resizebox{\textwidth}{!}{%
\begin{tabular}{cc|ccccccc}
       Metric & Parameter  & DL-Prior + Sampling  & DL & DL Ensemble & LMI & Edema Center of Mass & CET Center of Mass \\
\hline
 &Origin [mm]     & \textbf{3.51 $\pm$ 0.19} &3.72 $\pm$ 0.20  & 3.57$\pm$ 0.19 & 8.72 $\pm$ 0.27 *& 15.87  $\pm$ 0.73 * & 12.66 $\pm$ 0.56 * &  \\
MD &$\mu_D$ [mm]  & -0.30$\pm$ 0.08 & \textbf{-0.24$\pm$ 0.06}  & -0.25$\pm$ 0.06 & 1.66$\pm$ 0.15 * && \\
 &$\mu_\rho$      & \textbf{0.01 $\pm$ 0.02} & -0.02 $\pm$ 0.02& 0.05 $\pm$ 0.02& 0.01 $\pm$ 0.06 && \\
 
\hline
 &Origin [mm]    & 4.38 $\pm$ 0.24 & 4.56 $\pm$ 0.24   & \textbf{4.37 $\pm$ 0.24} & 10.30 $\pm$ 0.31 *& 18.64 $\pm$ 0.88 *& 14.9$ \pm$ 0.96 *\\
RMSD &$\mu_D$ [mm]   & 1.05 $\pm$ 0.10 & 0.84 $\pm$ 0.06 & \textbf{0.82 $\pm$ 0.06} & 2.56 $\pm$  0.13 * & &\\
 &$\mu_\rho$        & 0.40 $\pm$ 0.06 & 0.33 $\pm$ 0.03& \textbf{0.33 $\pm$ 0.03}& 0.74 $\pm$ 0.07 & &\\

\end{tabular}}
\end{table*}

\subsubsection{Estimated Tumor Concentration}
Due to the variations in the employed methodologies, exclusive reliance on parameter estimation is inadequate. A comprehensive evaluation of the resulting tumor concentration is imperative, especially considering the intended use for individualized radiotherapy. Therefore, we compare the MAE, the MSE, and the Dice-score for different threshold values. 

The errors of the presented methods are shown in Table \ref{table:concentraionResults}. It is found that our proposed in-patient-space DL ensemble and single networks (DL) perform significantly better than the atlas-registration-based LMI. 

\begin{table}
\centering
\caption{The mean absolute (MAE) and mean squared error (MSE) are compared between the predicted tumor concentration and the ground truth for different algorithms. A significant improvement of DL-Prior + sampling over the second-best method 
is demonstrated.}
\label{table:concentraionResults}
\resizebox{\columnwidth}{!}{%
\begin{tabular}{c|cccc}
    & MAE [$10^{-4}$] & MSE [$10^{-5}$] & Dice 0.5\\
\hline
DL-Prior + Sampling & \textbf{16 $\pm$ 2} & \textbf{30 $\pm$  7} & \textbf{0.93 $\pm$ 0.01}  \\
DL & 24 $\pm$  2 * & 68 $\pm$  10 * & 0.88 $\pm$ 0.01 *\\
DL Ensemble & 27 $\pm$  3 * & 72 $\pm $ 11 * & 0.89 $\pm$ 0.01* \\
LMI & 79 $\pm  6 $ * & 401$\pm$ 40 * & 0.75 $\pm$ 0.01 *\\
\end{tabular}}
\end{table}

\begin{figure}[H]
    \centering
    \includegraphics[width=\columnwidth]{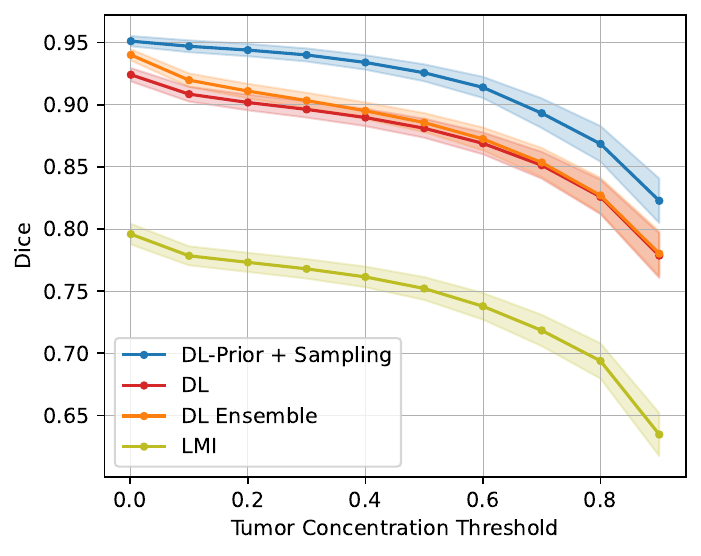}
    \caption{Mean Dice-score results between prediction and ground truth at different threshold concentrations. The single network approach is compared to the ensemble mean, LMI and the final combined version with CMA-ES sampling + DL-Prior.}
    \label{fig:diceResults}
\end{figure}

Additionally, we compare the Dice-score between thresholded ground truth and predicted tumor cell distributions for a clinically more relevant measure. Figure \ref{fig:diceResults} shows a general decrease in the Dice-score for increasing thresholds independent of the applied method. This decrease is intrinsic to the Dice, which is biased towards larger volumes. 

Compared to the atlas-based DL approach (LMI), there is a significant improvement in the Dice-score (Figure \ref{fig:diceResults}) for all of the shown methods. The single network already reaches Dice-scores of about 90\% in the clinically relevant range of low tumor concentrations. This result is slightly, but not significantly, outperformed by the ensemble mean. Still, the ensemble provides an additional variance utilized by the subsequent sampling approach in the form of a prior.

\subsection{Subsequent Evolutionary Sampling}
In the second part of the results, we focus on subsequent sampling methods in combination with the DL-Prior, as described in Figure \ref{fig:overview} c.

\subsubsection{Qualitative Results}
The combined method of DL-Prior and CMA-ES sampling (Figure \ref{fig:overview} e) closely aligns with the ground truth (Figure \ref{fig:overview} a). These qualitative findings are validated with quantitative results in the following sections.

\subsubsection{Estimated Parameter Comparison}
Similar results to DL and DL ensemble are found for subsequent evolutionary sampling (Table \ref{table:parameterResults}). Limited information, introduced by stochastic thresholding of test data, might lead to a performance threshold that is impossible to overcome theoretically. For the combined version of sampling and prior, a slightly worse result is found for $\mu_D$ and $\mu_\rho$, whereas the origin deviation is comparable to the ensemble mean and significantly better than the center of mass.

Even though parameter estimation does not improve with the network ensemble and subsequent evolutionary sampling, the rationale for training the ensemble is founded on the anticipation of achieving more stable predictions and providing variance estimation for the DL-Prior to optimally initialize the subsequent sampling. Further, the ill-posedness of the problem allows multiple parameter combinations to result in similar tumors. The following section examines this premise by comparing the resulting tumor concentrations.

\subsubsection{Estimated Tumor Concentration}
\label{sec:res_tumorConc}
Subsequent sampling decreases the error to the ground truth of the estimated tumor concentration as shown in Table \ref{table:concentraionResults}. This highlights the synergistic combination of sampling with a DL-Prior. Further, the combined version of CMA-ES sampling and DL-Prior achieves a significantly better Dice-score than the DL-Prior alone over the whole range of tumor concentrations (Figure \ref{fig:diceResults}). Within the clinically relevant range, up to 40\% tumor concentration, a Dice overlap of 95\% is found, further demonstrating the superiority of combining DL with traditional sampling.

\subsubsection{Runtime comparison}
\begin{figure}
    \centering
    \includegraphics[width=\columnwidth]{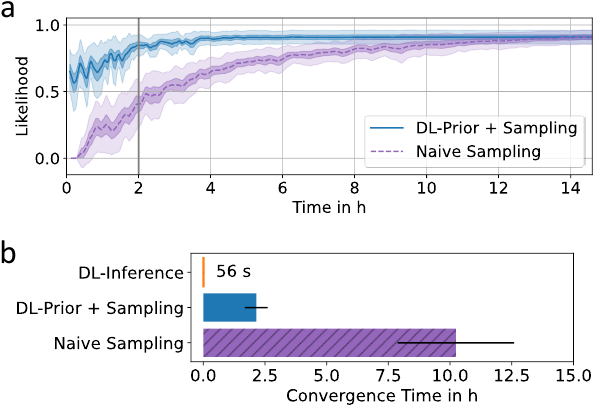}
    \caption{(a) The mean likelihood over time for Naive Sampling and DL-Prior + Sampling is shown. Shaded areas represent standard error over different patients. DL-Prior + Sampling consistently yields higher likelihoods, with a significant difference around the clinically relevant 2-hour mark, resulting in a likelihood of 0.85 $\pm$ 0.03, while Naive Sampling only reached 0.41 $\pm$ 0.05. This translates to a p-value of \(1.2 \times 10^{-5}\) and an effect size of 2.0. (b) The mean convergence time of our DL-inference (orange) and our combination of classical sampling with DL-Prior (blue) is compared to sampling without prior (purple, dashed). The paired t-test results in an effect size of 1.0 and a p-value of 0.003}
    \label{fig:convTime}
\end{figure}

In Figure \ref{fig:convTime}, we compare the convergence times across various methods for a subset of samples. Convergence time for the evolutionary sampling method is defined as the duration required to achieve 99\% of the maximum posterior probability. All calculations were performed on a single CPU to maintain consistency in comparison. The convergence time associated with evolutionary sampling using a DL-Prior is not directly comparable to the inference time of DL alone, as the latter is predominantly determined by the duration of a single forward simulation run, while the duration of the DL inference is negligible. When comparing against naive evolutionary sampling, integrating a DL-Prior into our optimization strategy demonstrates an 80\% improvement in convergence time.

To guarantee the integrity of our comparative analysis regarding time, we have included performance metrics for the specific subset of samples utilized in the temporal comparison, as detailed in Table \ref{table:timeComparisonPerformance}. Sampling enhanced with a DL-Prior demonstrates improved results in terms of the Dice. However, the observed differences between these methods fall within the range of the method-specific error margins. These observations suggest that both approaches converge towards a similar minimum, legitimizing the time comparison of Figure \ref{fig:convTime}.

\begin{table}[H]
\centering
\caption{Tumor concentration comparison between DL-Prior with subsequent sampling and sampling without a prior for a subset of patients. The Dice-score at different thresholds (0.01, 0.1, 0.5) is given, and no substantial difference was found. Simulation time differences are shown in Figure \ref{fig:convTime}. }
\label{table:timeComparisonPerformance}

\resizebox{\columnwidth}{!}{%
\begin{tabular}{c|cccc}
      Metric & DL-Prior + Sampling & Naive Sampling \\
\hline 
Dice at 0.01 & \textbf{0.94} $\pm$ 0.02 & 0.90 $\pm$ 0.02 \\
Dice at 0.1  & \textbf{0.94} $\pm$ 0.03 & 0.92 $\pm$ 0.03 \\
Dice at 0.5 &  0.93 $\pm$ 0.02 & \textbf{0.95} $\pm$ 0.03 
\label{tab:WithWithoutProir}
\end{tabular}}
\end{table}

\section{Validation on Patient Data}
\begin{figure}[h!]
    \centering
    \includegraphics[width=\columnwidth]{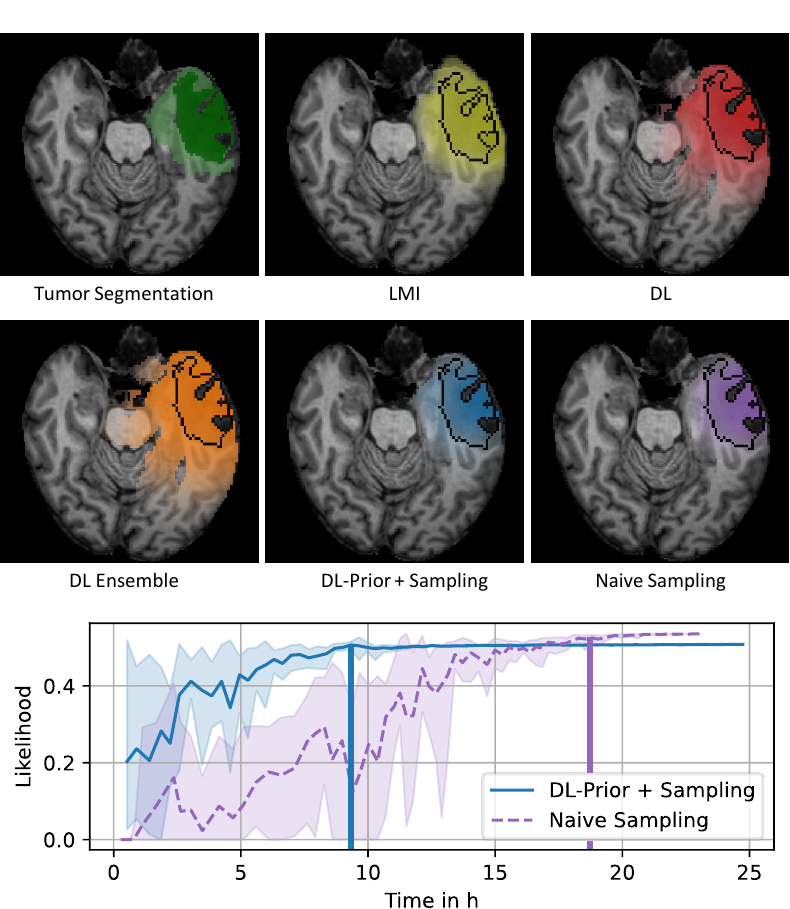}
    \caption{Real patient example with different inversion strategies. The first image shows the ground truth tumor segmentation before surgery. Light green shows the edema region, while dark green represents the CET, which is included in the following images as a black outline. The mean likelihood development over time comparing DL-Prior sampling to naive sampling is plotted. The margin shows the minimum and maximum values for each CMA-ES epoch, and the vertical lines depict the time points of convergence. The likelihood for the other methods is 0.39 for LMI, 0.23 for DL, and 0.13 for DL ensemble.}
    \label{fig:realPatients}
\end{figure}

To demonstrate that simulation advancements can translate into clinical application, we tested our method on nine patients. An example is shown in Figure \ref{fig:realPatients}. As already found in Figure \ref{fig:example}, the LMI simulation is not limited to white and gray matter regions and, therefore, grows into the CSF region due to registration inaccuracies. DL and DL ensemble already roughly estimate the tumor location, while overestimating the volume, while DL-Prior + sampling and naive sampling estimate the segmentation well. The faster convergence becomes evident, considering the likelihood over time. 

\begin{figure*}[h!]
    \centering
    \includegraphics[width=\textwidth]{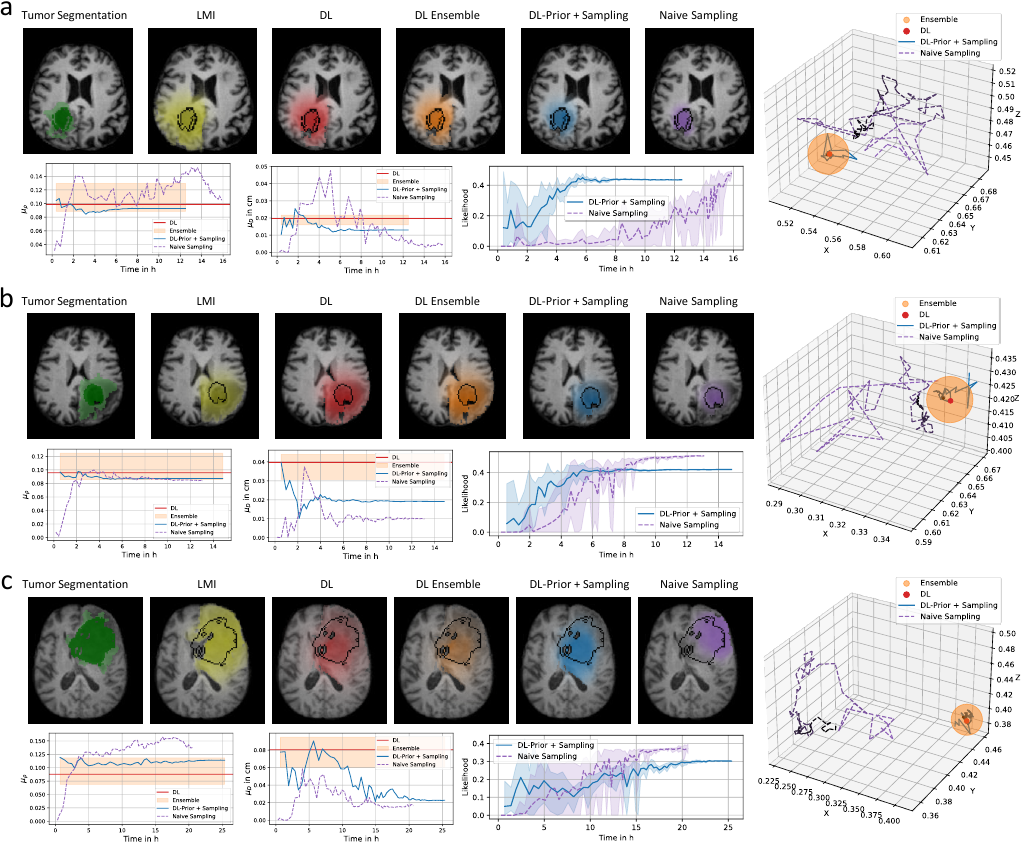}
    \caption{Evaluation of our method focusing on multiple scans of varying sizes and mass effects, particularly analyzing challenging cases. The final tumor concentration predictions are shown in the first row. Below, the time developments of $\mu_D$, $\mu_\rho$, the likelihood, and the tumor origin are shown. The DL and ensemble results are shown in red and orange. It becomes apparent that for small tumors without mass effect (a), the DL-Prior estimation provides a solid estimate leading to faster convergence. Large mass effects (c) can result in a failure of the network, constraining the sampling process into a suboptimal parameter set, which becomes apparent concerning the origin in this case.}
    \label{fig:moreExamples}
\end{figure*}

The quantitative ability of different methods to fit real tumor segmentations is compared in Table \ref{tab:realPatients}. The Dice-score is compared for CET and edema regions. It is clearly visible that sampling improves the simulation fit dramatically. The differences in mean Dice-score between naive sampling and prior included sampling are small, but those results do not allow for statistical statements. Therefore, we qualitatively analyze the results case-specific in Figure \ref{fig:moreExamples}. Regarding the runtime, a clear advantage is seen for DL-Prior + sampling.

\begin{table}[H]
\centering
\caption{For nine real patients, the Dice-score to the tumor segmentation is compared for different algorithms, thresholded at the $\overline {c^{th}} = \{ 0.25, 0.675 \} $. Additionally, the convergence time is reported for each method. It becomes apparent that naive sampling and sampling with a prior reach similar Dice-score with significant time differences in runtime.}
\label{tab:realPatients}
\resizebox{\columnwidth}{!}{%
\begin{tabular}{c|cccc}
    & Dice CET & Dice Edema & Runtime\\
\hline
DL-Prior + Sampling & 0.68 $\pm$ 0.03 & 0.68 $\pm$ 0.02 & (4.9 $\pm$ 1.2) h \\
Naive Sampling      & 0.70 $\pm$ 0.03 & 0.70 $\pm$ 0.02 & (16.3 $\pm$ 1.7) h*\\
DL                  & 0.39 $\pm$ 0.06* & 0.44 $\pm$ 0.04* & $<$ 1 min*\\
DL Ensemble         & 0.44 $\pm$ 0.08* & 0.47 $\pm$ 0.05* & $<$ 1 min*\\
LMI                 & 0.34 $\pm$ 0.09* & 0.55 $\pm$ 0.04* & $<$ 1 min*\\
\end{tabular}}
\end{table}

Our method generally demonstrates strong performance, as illustrated in Figure \ref{fig:convTime}, which also shows significant runtime improvements. However, there are occasional instances where it encounters challenges. We analyze various example scans in Figure \ref{fig:moreExamples} to assess the types of tumors that our method accurately characterizes and to identify specific scenarios where our approach falls short, highlighting potential for future research.
For the small tumor (a), the likelihood plot indicates a strong advantage for including the DL-Prior convergence, which is reached much faster than with naive sampling. The naive sampling approaches the optimal prior region only after exploring a larger space, while the diffusion and growth rate stay close to the initial prior.
For the larger tumor (b), the prior estimate of the origin is close to the final estimate of the naive process. The prior sampling early advantage diminishes over time, as the DL-Prior restricts the diffusion rate to decrease further, which seems beneficial for this tumor.
However, for the example with large mass effect (c), all methods struggle to fit the tumor, which is expected, as the mass effect is not part of the physical tumor model. Due to the insufficient tumor model, the DL-Prior becomes obsolete, and no improvement in likelihood is achieved. Neither method fits the scans well in this case.

In summary, the introduction of a DL-Prior helps the sampling strategy to converge faster to the optimal solution (Figure \ref{fig:convTime}, \ref{fig:moreExamples} a). In some cases, the underlying tumor model is clearly oversimplified (\ref{fig:moreExamples} b), and in a few other cases, the prior can lead the sampling in the wrong direction (\ref{fig:moreExamples} c). The last problem could be overcome by relaxing the prior constraint over time.

\section{Discussion}

We developed a novel technique combining DL with CMA-ES sampling to estimate brain tumor density based on magnetic resonance images, synergistically exploiting the unique advantages of each approach. 

A new DL-based method is introduced, operating solely in patient image space compared to previous atlas-based methods \cite{ezhov2023learn} by explicitly incorporating patient anatomy into the network. With our method, a better parameter estimation is found (Table \ref{table:parameterResults}). Additionally, a better Dice overlap to the ground truth concentration is observed (Figure \ref{fig:diceResults}). We assume that the abundance of registration errors between the patient and the atlas space causes this improvement. 

No significant improvement in parameter estimation is achieved by combining multiple fine-tuned networks into an ensemble, whereas a significant improvement in Dice overlap is found compared to the single network. This can be attributed to the fact that the network ensemble provides higher stability in prediction \cite{ju2018relative}.

The primary strength of the proposed method lies in the combination of DL and CMA-ES sampling. Within the confines of our simulation, we observed that the combined method provides a faster and more accurate representation of tumor dynamics compared to individual approaches (Figure \ref{fig:diceResults}).

In synthetic data, the combined optimization time was improved by a factor of five compared to the optimization without prior (Figure \ref{fig:convTime}) while the performance is increased to 95\% Dice compared to 80\% reached by a previous DL approach \cite{ezhov2023learn} (Figure \ref{fig:diceResults}).

For real data, the advantage of sampling over pure DL methods becomes apparent. The DL-Prior reduced the convergence time by a factor of three with similar results.

Regarding brain tumor modeling in general, despite being enclosed inside the skull, it is highly debatable to which degree the complex biological system of brain tumors can be modeled using deterministic models, to which degree it results from chaotic behavior, and what role intratumoral heterogeneity and evolution play. Despite this, modeling-based approaches open a unique perspective on the diffuse spread of tumor cells, which might be sufficient for radiotherapy improvement.

Regarding the validity of our findings, the model's limitations are primarily centered around its simplicity, which, while beneficial for understanding the proposed combination of two modeling worlds, may not accurately represent the underlying complex biological systems. The model's exclusion of necrosis and mass effect factors is a significant limitation. Additionally, the isotropic diffusion, particularly the distinction between white and gray matter, is somewhat arbitrary, potentially oversimplifying these complex dynamics. Furthermore, the main findings of the method are confined within the narrow boundaries of these simulations, where tumor behavior is strictly deterministic. Therefore, extended real data validation with advanced tumor models is critical to validate our approach.

Regarding clinical implications, we demonstrate a method to combine general, flexible sampling methods with specialized DL. Changing the DL model, e.g., with respect to higher tumor model complexity, can be time-consuming since the network must be retrained. However, this should not be an issue in clinical use, as the model will not be updated frequently.

\section{Outlook}
For an impact on radiotherapy planning, more sophisticated tumor growth models are needed to explain the data. 
The presented combination of DL and classical sampling enables more complex models and models with more parameters to be solved in clinically relevant time frames. Such models are crucial for several well-studied tumor properties like mass effect \cite{lipkova2022modelling, subramanian2020multiatlas}, nutrient flow \cite{frieboes2006integrated}, and necrosis \cite{zheng2005nonlinear}. The mass effect takes the tumor-induced tissue shift into account, whereas nutrient flow and necrosis formation affect the growth dynamic of the tumor based on complex interaction with itself and surrounding tissue.

It has been shown that tumors grow along fiber tracts, explaining the so-called butterfly tumors \cite{finneran2020long, siddiqui2018butterfly}. Additional imaging modalities like diffusion tensor imaging contain information about fiber tracts that can be incorporated in the tumor model \cite{hathout20163, painter2013mathematical, van2023evaluating}. Therefore, the shown framework can be adapted and finetuned in multiple ways. For example, by incorporating multiple simulation time points into the CMA-ES selection process or by relaxing the prior weight over time. 

Finally, an evaluation of large-scale clinical data is needed. As the availability of longitudinal data is highly limited due to the rapid surgery after detection for most patients, other validation metrics are considered. Prediction of tumor recurrence represents one of the most promising approaches. Besides the influence of treatment procedures like surgery and radiotherapy, it was shown that a certain degree of recurrence prediction is possible \cite{metz2020predicting, lundemann2019feasibility, metz2024toward}. Showing the ability to predict recurrence with a physically constrained model would provide an excellent argument for proceeding with clinical trials testing model-based radiotherapy planning.

\section{Conclusion}

This study introduces a novel framework that synergistically combines deep learning with evolutionary sampling to enhance brain tumor modeling. By leveraging a DL ensemble for initial parameter estimation and incorporating it into a high-precision evolution strategy, we achieve a fivefold acceleration in convergence time and a Dice-score of 95\%. This approach significantly constrains the sampling parameter space, offering a powerful and efficient solution for individualized radiotherapy planning with the potential to substantially improve patient outcomes.

\bibliography{bib.bib}{}
\bibliographystyle{IEEEtran}

\end{document}